\documentstyle[floats,prl,aps,epsf]{revtex}
\draft 
\begin{document}

%
%
\input epsf
\renewcommand{\topfraction}{0.8}

\twocolumn[\hsize\textwidth\columnwidth\hsize\csname
@twocolumnfalse\endcsname
%

\def\lesssim{\mathrel{\hbox{\rlap{\hbox
{\lower4pt\hbox{$\sim$}}}\hbox{$<$}}}}
 
\def\gtrsim{\mathrel{\hbox{\rlap{\hbox
{\lower4pt\hbox{$\sim$}}}\hbox{$>$}}}}

\title{Three-dimensional adaptive evolution of gravitational
waves in numerical relativity}
\author{Kimberly C. B. New,$^{\rm 1,2} $Dae-Il Choi,$^{\rm 1}$
Joan M. Centrella,$^{\rm 1}$ Peter MacNeice,$^{\rm 1,3}$
Mijan F. Huq,$^{\rm 4}$ and Kevin Olson$^{\rm 5}$}
\address{$^{\rm 1}$Department of Physics, 
Drexel University, Philadelphia, PA 19104}
\address{$^{\rm 2}$X-2, MS B-220,Los Alamos National Laboratory,
Los Alamos, NM 87545}
\address{$^{\rm 3}$Code 931, NASA/GSFC, Greenbelt, MD 20771}
\address{$^{\rm 4}$Dept. of Astronomy \&  Astrophysics, and
Center for Gravitational Physics \& Geometry,}
\address{The Pennsylvania State University, University Park, PA 16802}
\address{$^{\rm 5}$Enrico Fermi Institute,University of Chicago,
Chicago, IL 60637}


\maketitle

    \begin{abstract}
Adaptive techniques are crucial for successful numerical modeling 
of gravitational waves from astrophysical sources such as coalescing
compact binaries, since the radiation typically
has wavelengths much larger than the scale of the sources.  We
have carried out an important step toward this goal, 
the evolution of weak gravitational waves using adaptive mesh
refinement in the Einstein equations.  The 2-level adaptive
simulation is compared with unigrid runs at coarse and fine
resolution, and is shown to track closely the features of the fine grid
run.
    \end{abstract} 


\vskip2pc]

A new era in general relativistic research is beginning with the
inauguration of a worldwide network of gravitational wave
observatories including LIGO, VIRGO, GEO, and TAMA \cite{detectors}.
The signals that these instruments will detect are expected to 
often arise in highly
nonlinear, dynamical astrophysical events, such as the final 
merger of coalescing
compact
binaries.  Fully general relativistic simulations of these events 
in three-dimensions
will be an
essential component
of the successful detection and interpretation of gravitational 
wave signals.

A key difficulty in calculating accurate waveforms from these
simulations lies in the fact that the models encompass both length and
time scales that can vary by an order of magnitude or more.  Adequate
resolution is needed to model the strong-field dynamics near the
sources.  Also, the computational
domain must be sufficiently large and well-resolved to model the
resulting gravitational radiation that, for coalescing compact
binaries, can have wavelengths $\sim 10 - 100$ times the size of the
sources.  Adaptive techniques that offer dynamic variable 
resolution are thus crucial for success.

Adaptive mesh refinement (AMR) has undergone significant 
developments during the past two decades, and remains an active
area of research and development.
 Within the context of general relativistic simulations,
 Choptuik first implemented the Berger-Oliger \cite{BO} AMR
method to study critical phenomena in the collapse of 
massless scalar
fields in spherical symmetry \cite{choptuik1,choptuik2}; see
also \cite{hamade}. AMR techniques were applied to spherical black
hole evolutions in Refs.\ \cite{masso} and \cite{wild_thesis}.
  In addition, the construction of initial data for black
hole collisions using AMR was studied in Ref.\ \cite{DJKN}, and
fixed mesh refinement was employed for a short part of a binary
black hole calculation in Ref.\ \cite{brugmann}.  Recently,
Papadopoulos, Seidel and Wild \cite{PSW} carried out 3-D
adaptive computation of gravitational waves using a single
model equation that describes perturbations of a non-rotating
black hole.

We have carried out a 3-D simulation of pure gravitational 
waves in the Einstein equations using AMR.  While our numerical
relativity code does solve the full Einstein equations, we have
chosen to evolve linearized Teukolsky waves to enable comparison
with an analytic solution for this simulation.
  Overall, our 2-level adaptive 
simulation successfully reproduces the features of a uniform
grid run at the finer resolution.

Our numerical relativity code is based on the ADM code
developed by the BBH Alliance \cite{BBHcode} and
is written in Cartesian coordinates.  We have implemented a
conformal ADM formalism \cite{SN95,BS99}, and  have 
incorporated  octant symmetry boundary conditions
to minimize the number of gridpoints.  
Parallelism and AMR have been implemented using
Paramesh \cite{paramesh}, 
as discussed below.  The 
time evolution is carried out using an 
iterative Crank-Nicholson method with two iterations
\cite{ST99}. 
Interpolated Sommerfeld outgoing wave conditions
are applied at the outer 
boundary of the grid \cite{SN95,BS99}.

Given the complexity of this system of equations and the computational
techniques employed, analytic solutions provide an essential testbed
for the resulting numerical code.  To this end, we use the
Teukolsky wave \cite{ST82}, which is a time-dependent solution to 
the linearized vacuum Einstein equations with 
geodesic slicing, $\alpha = 1$  and $\beta = 0$.  It is based on a
generating function 
$F(t,r) = (A/\omega^2) (t \pm r) \exp(- (t \pm r)^2/\omega^2)$,
where $A << 1$ is an amplitude parameter and $\omega$ is the
width of the generating function, which is of the same 
order as the  wavelength
of the gravitational wave.  To set initial data for our simulations,
 we use a time symmetric,
even-parity $L=2$, $M=0$
Teukolsky wave solution that contains
a combination of ingoing and outgoing gravitational waves; see Refs. 
\cite{ST82,Eppley}.  We set $\omega = 1$ and take $A = 10^{-6}$,
so that the metric perturbation has an initial amplitude 
$g_{ij} - 1 \sim \pm 10^{-5}$.  Since
$t=0$ is a moment of time symmetry, we have $K_{ij}$ = 0
initially.  

This data is then evolved in our code 
with $\alpha =1$ and $\beta=0$.
The outgoing waves travel directly toward the
edge of the grid.  The initially ingoing waves first travel toward
the origin, then reflect and move outward.  As the overall signal
travels away from the origin, it leaves flat space behind. 
Ref. \cite{ST82} gives explicit 
expressions for the metric components $g_{ij}$ that we use
as a first-order analytic solution for comparison with our numerical
results. All simulations presented in this paper were done with 
octant symmetry.

We employ an AMR scheme that works on logically Cartesian, or
structured, grids.  The pioneers in this area are Berger and
co-workers \cite{BO,berger_phd,BC}, who covered the computational
domain with structured grids and sub-grids that could overlap,
take arbitrary shapes, be rotated with respect to the coordinate
axes, and be merged with other sub-grids at the same level of
refinement.  Although this strategy is flexible and memory-efficient,
the resulting code can be very complex and difficult to parallelize,
especially in 2D and 3D.  Simplifications of this basic scheme were
developed by Quirk \cite{quirk} and later by DeZeeuw and Powell
\cite{deZP}, who implemented refinement by bisecting the appropriate
grid blocks in each coordinate direction and linking the hierarchy
of sub-grids as the nodes of a data tree.  Wild and Schutz 
\cite{wild_thesis,wild-schutz} have developed a somewhat different
approach using hierarchical linked-lists. 

Paramesh uses an AMR technique similar to that of
DeZeeuw and Powell \cite{deZP} in which grid blocks are bisected in
each coordinate direction when refinement is needed.  The grid
blocks all have the same logical structure,
with $nxb$ zones in the
$x-$direction, and similarly for $nyb$ and $nzb$.
Thus, in 3D refinement of a block yields
8 child blocks, each having $nxb \times nyb \times nzb$ zones but with 
 zone sizes a factor of two smaller than in the parent block.
Refinement can then continue on any or all of these child blocks,
with the restriction that the grid spacing is not allowed to change
by more than a factor of two, or one refinement level, at any location
in the spatial domain.  Every grid block has a number of guard cell
layers along each of its boundaries, to allow operations between
neighboring blocks as well as 
between different refinement levels. 
 These guard cells are filled
with data from neighboring blocks or, if the block is along the
edge of the computational domain, with appropriate outer boundary
conditions. 

Paramesh handles the creation of grid blocks and builds and 
maintains the data structure (quad-tree in 2D, oct-tree in 3D)
that tracks the spatial relationships between blocks.  It also
handles all inter-block communications.  In addition, it 
keeps track of 
physical boundaries on which particular conditions are to be set,
guaranteeing that the child blocks inherit this information from the
parent blocks.  
When executing on a parallel machine, Paramesh distributes the blocks 
among the available processors to achieve load balance,
maximize block locality, and  minimize inter-processor
communication.

We have implemented our numerical relativity code in the
Paramesh framework.  For the runs presented in this paper,
we use blocks with $nxb = nyb = nzb \equiv nb$ and grid sizes
$\Delta x = \Delta y = \Delta z \equiv h$.  We also use the 
same timestep, chosen to give stability on the finest
grid, over the whole computational domain.  The basic
variables are located at the centers of the grid cells.
We use second-order spatial finite 
differences and a single layer of guard cells. For the AMR
runs, we use the restriction (transfer of data from fine to
coarse grids) and prolongation (coarse to fine) operators
built into Paramesh, which employ linear interpolation.  All the 
calculations presented in this paper were done using a T3E.

A necessary preliminary step in using AMR for numerical simulations
is to verify that the code works correctly when run using a single
grid with constant $h$ that covers the entire computational
domain, {\em i.e.}
for unigrid simulations.
Of particular importance is the convergence behavior of the code.
 Since we are using finite differences that
are second-order accurate in both space and time, we expect that the
errors will decrease $\sim {\cal O}(h^2)$ as $h \rightarrow 0$.

Using the Teukolsky wave initial data, we ran a set of three
runs with increasingly finer grids:  $h = 0.25$ ($16^3$ grid),
$h = 0.125$ ($32^3$ grid), and $h = 0.0625$ ($64^3$ grid).  In
these runs, the outer boundaries of the grid were placed at
$x_{\rm max} = y_{\rm max} = z_{\rm max} = 4$.
Calculation of various error norms demonstrated that, even for these
relatively modest resolutions, the code has nearly second-order
convergence for the propagation of the wave across the grid.
However, as the waves move off the outer
edge of the grid, lower amplitude reflected waves travel
inward from the outer boundary.
When the outer boundary is moved farther from the origin, and thus
closer to the asymptotic region in which the Sommerfeld outgoing wave
conditions are applicable, the errors due to boundary effects
decrease. 

To evolve gravitational waves using
AMR, we 
have chosen to use a refinement criterion based on the
first derivative of the diagonal components of the conformal
metric $\tilde{g}_{ij} = e^{-4 \phi} g_{ij}$, where 
$e^{4 \phi} = {\rm det}(g_{ij})^{1/3}$.
Thus, in each zone we calculate the quantity
$\tau  =  \frac{h}{f}
 \left( {\partial f \over \partial x} 
      + {\partial f \over \partial y} 
      + {\partial f \over \partial z} \right ),$
where $f = \tilde{g}_{ij}-1$ for $i = j$. 
The computational domain is covered with blocks having
$nb^3$ physical zones.
 A block is marked for
refinement if $\tau > \tau_{\rm re}$ for at least one zone in that
block, and for derefinement if $\tau < \tau_{\rm de}$ for all 
zones in that block, where $\tau_{re} \ge \tau_{de}$.  Note that
both refinement and derefinement in Paramesh are restricted by the
requirement that the grid can change by only one refinement level
at block boundaries.

We concentrate on 2-level AMR, with level 1 indicating the coarse
grid and level 2 the fine grid.  
For the simulation presented here, we
 use grid blocks with $nb =4$ and
set the outer boundary of the computational
domain at $x_{\rm max} = y_{\rm max}
 = z_{\rm max} = 7$.  After some experimentation, we chose to trigger
refinement by  
$\tau_{re} = 1.75$, and derefinement by $\tau_{de} = 0.875$.

To set up the initial data, we first covered the entire computational
domain with level 1 blocks, and used the Teukolsky solution to
give $\tilde g_{ij}$ and $\phi$ .  The
quantity $\tau$ was then computed and
the refinement criterion applied, yielding blocks at level 2.
The Teukolsky solution was again applied to these refined blocks
to set up $\tilde g_{ij}$ and $\phi$.  However, when the first
derivatives of $\tilde g_{ij}$ were taken numerically to compute
the conformal
connection functions $\tilde \Gamma^i$ \cite{BS99}, 
discontinuities in $\tilde \Gamma^i$ appeared
at the boundaries between level 1 and level 2 blocks.  When this
data was then evolved forward in time, these initial discontinuities
produced errors in other variables.  To
circumvent these problems, we calculated
$\tilde \Gamma^i$ on a level 2 grid covering the whole domain,
and then used the restriction operator to give $\tilde \Gamma^i$ on the
level 1 blocks.  This produced a data set without initial
discontinuities.

\begin{figure}
\centerline{\epsfxsize=12cm\epsfysize=12cm\epsffile{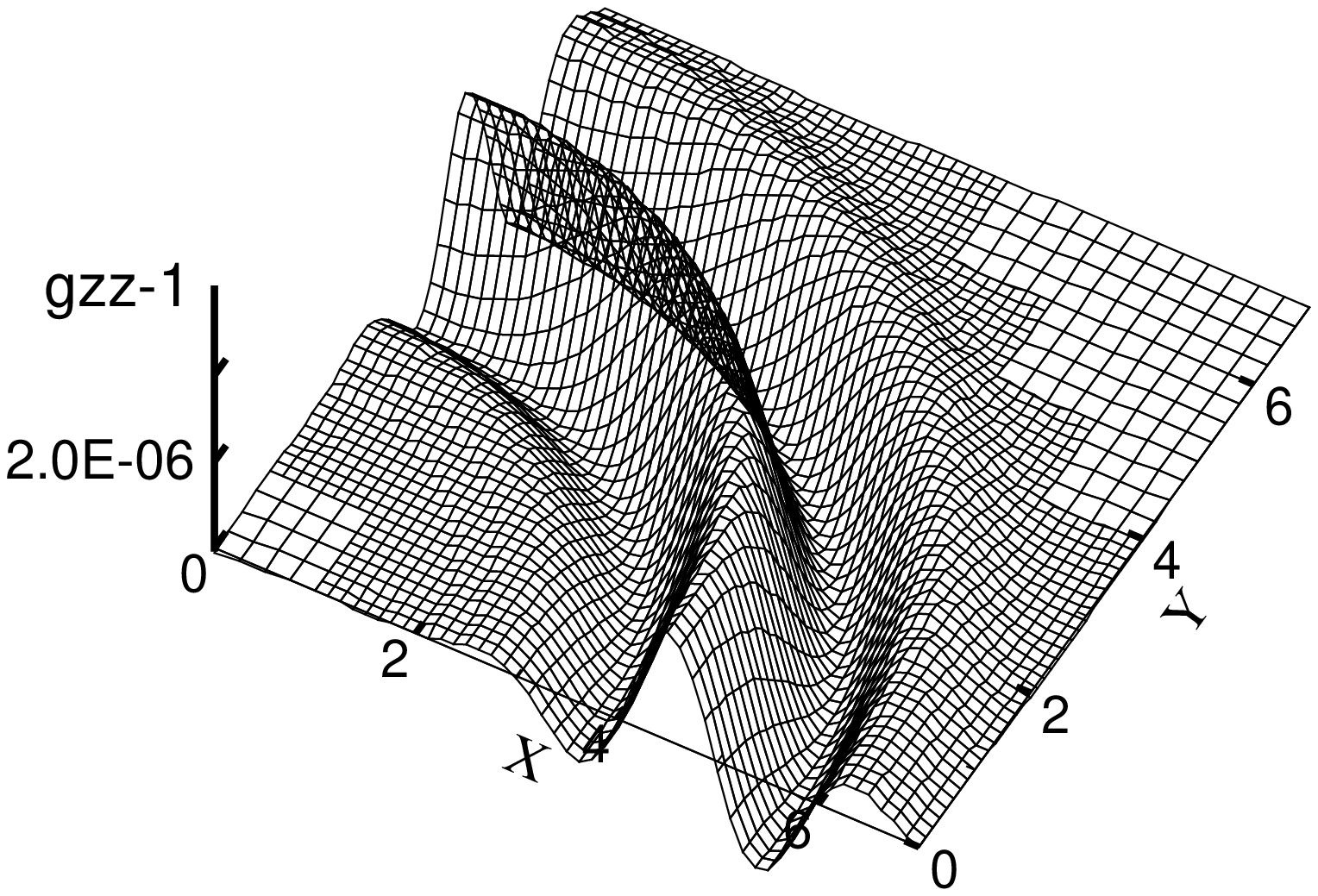}}       
\caption{The metric component $g_{zz}$ is shown in the equatorial
plane
at time $t = 5.06$  in the 2-level AMR run.
}
\label{gzz_2D}
\end{figure}

Figure~\ref{gzz_2D} shows the metric component
$g_{zz}$ in the equatorial plane for an AMR run with resolution
$h = 0.25$ on level 1 and $h = 0.125$ on level 2 at
time $t=5.06$.   Initially, only
the region near the origin is covered by the fine grid.  This 
refined region moves outward as the wave propagates toward the
edge of the grid.  As seen in Fig.~\ref{gzz_2D},
the region  near the origin is
derefined as the solution approaches flat space in that region.

\begin{figure}
\centerline{\epsfxsize=12cm\epsfysize=12cm\epsffile{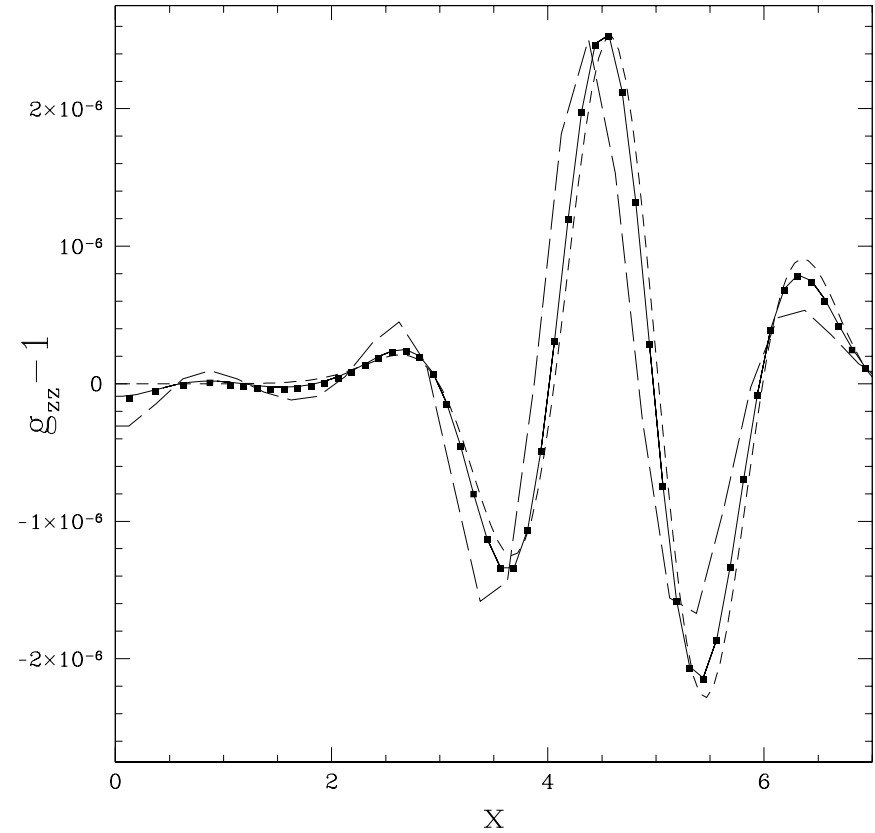}}       
\caption{The quantity $g_{zz} - 1$ is shown along the $x-$axis at
time $t = 5.06$.  The filled squares show the data for the 2-level AMR
run, the short dashed line the analytic result,
 the solid line the unigrid result
with $h = 0.0625$, and the long dashed line the unigrid result with
$h = 0.125$.
}
\label{gzz_1D}
\end{figure}

It is interesting to compare this AMR
solution with unigrid runs using the same computational domain.
  Figure~\ref{gzz_1D} shows 
the quantity $g_{zz} -1$ along the $x-$axis  at $t=5.06$, with
the data from the AMR run shown as filled squares.
In addition, a short dashed line shows 
the analytic solution, a solid line 
the unigrid result with 
$h = 0.125$ ($56^3$ grid, same resolution as the level 2 AMR grid)
 and a long dashed line the unigrid result with
$h = 0.25$ ($28^3$ grid, same resolution as the level 1 AMR grid).
Notice that the AMR solution closely tracks the higher resolution
unigrid result. When we take the $L_1$, $L_2$, and $L_{\infty}$ error
norms of $g_{zz}$ over the entire computational volume, we find that
the AMR results closely track those of the higher
resolution unigrid run.

\begin{figure}
\centerline{\epsfxsize=12cm\epsfysize=12cm\epsffile{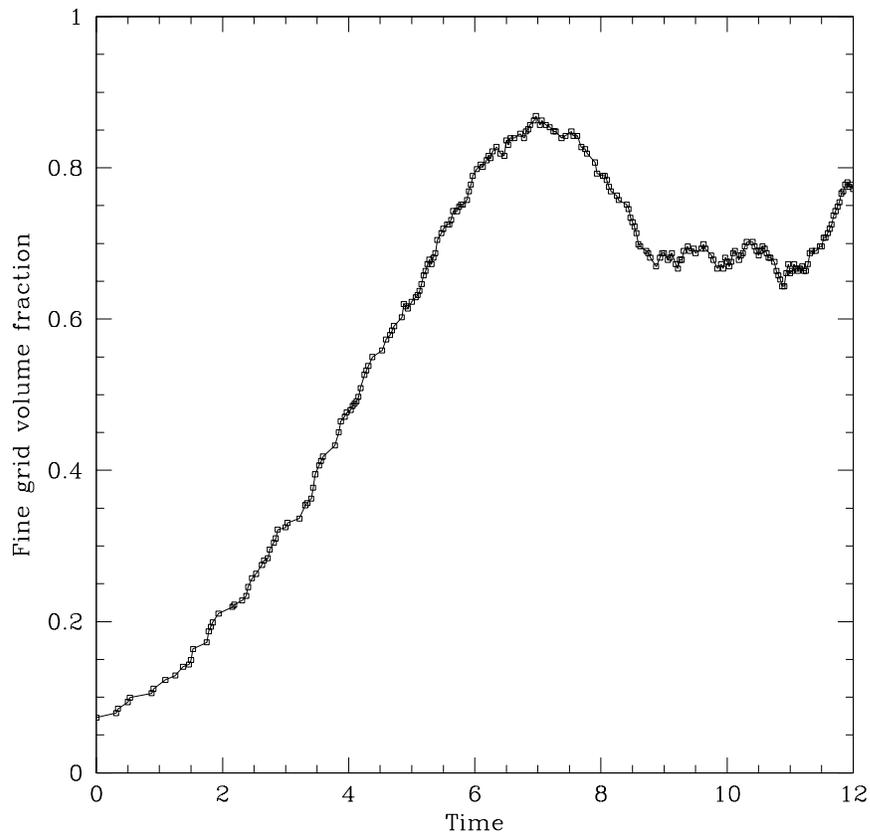}}       
\caption{The fraction of the computational volume covered by the fine
grid is shown versus time for the AMR run in Fig.~\protect{\ref{gzz_2D}}
}
\label{effic}
\end{figure}

Initially, less than $10\%$ of the computational volume was covered by
the fine grid.  This fine grid volume fraction increases as the wave
propagates outward, as shown in  Fig.~\ref{effic}. Although
derefinement in the equatorial plane first occurs around the origin at
$t \sim 5$, the total number of fine grid cells continues to increase
as the expanding wavefronts propagate outward.  The main peak of the
wave (located at $x \sim 4.5$ in Fig.\ref{gzz_1D}) encounters the outer
boundary at $t \sim 7.5$.  Around this time, Fig.~\ref{effic} shows
that the fine grid volume fraction begins decreasing as the derefined
region behind the wave grows.  By $t \sim 9$, most of the Teukolsky
wave has left the grid.  However, reflected waves traveling inward
from the outer boundary cause continued refinement and derefinement,
preventing the fine grid volume fraction from dropping to zero.

In this calculation, the savings in computational cost of the AMR
run over a unigrid run are modest, even if we neglect the problems 
with the boundary conditions, since the gravitational wavelength
$\lambda$ is comparable to the dimension $L$ of the computational
domain.  In a more realistic simulation, we would expect
$\lambda \ll L$, resulting in significant computational savings.
For a full simulation of binary coalescence, one will need to handle
multiple scales to evolve both the dynamics of the sources and the
propagation of the waves that develop.  While the sources are
expected to require high resolution, particularly near regions of
shocks or shear layers, the waves can likely be handled with
lower resolution.  Thus, although the resolutions used here are
modest compared to those in previous numerical studies of Teukolsky
waves, e.g. \cite{AMSST}, they are typical of the resolutions
likely to used for gravitational waves in full simulations of 
binary coalescence in the next few years.

While this AMR calculation consitutes an important step toward
the ultimate goal of realistic
models of gravitational wave sources, more work remains to
be done. For example,
even though
AMR can be used to alleviate
the problem of reflected waves
by moving the outer boundary even farther from the
origin, better outer boundary conditions are still needed \cite{SGBW}.
In addition, it will be interesting to test different
refinement criteria, including those based on truncation error
estimation \cite{choptuik2,PSW}.
Finally, we observed that 
 derefinement introduces some high frequency noise,
principally in the extrinsic curvature variables. While this did not
pose a significant problem for this 2-level run, it can become
important when more refinement levels are used.  We are
experimenting with various ways of removing this noise by using
filtering \cite{wild_thesis} or adding dissipation \cite{GARP,LT},
and will report on this work elsewhere.

We are pleased to thank Matt Choptuik  for interesting
and helpful discussions.
This work was supported in part by NSF grant
PHY-9722109 at Drexel, and  grant number PHY990002P  from the
Pittsburgh Supercomputing Center (PSC), which is 
supported by several federal agencies, the
Commonwealth of Pennsylvania and private industry.



\clearpage

\clearpage

\clearpage

\end{document}